# Domestic Activity Clustering from Audio via Depthwise Separable Convolutional Autoencoder Network

Yanxiong Li
*School of Electronic and Information Engineering, South China University of Technology, Guangzhou, China*
eeyxli@scut.edu.cn

Wenchang Cao
*School of Electronic and Information Engineering, South China University of Technology, Guangzhou, China*
wenchangcao98@163.com

Konstantinos Drossos
*Audio Research Group, Tampere University, Tampere,Finland*
konstantinos.drossos@tuni.fi

Tuomas Virtanen
*Audio Research Group, Tampere University, Tampere, Finland*
tuomas.virtanen@tuni.fi

*Abstract*—Automatic estimation of domestic activities from audio can be used to solve many problems, such as reducing the labor cost for nursing the elderly people. This study focuses on solving the problem of domestic activity clustering from audio. The target of domestic activity clustering is to cluster audio clips which belong to the same category of domestic activity into one cluster in an unsupervised way. In this paper, we propose a method of domestic activity clustering using a depthwise separable convolutional autoencoder network. In the proposed method, initial embeddings are learned by the depthwise separable convolutional autoencoder, and a clustering-oriented loss is designed to jointly optimize embedding refinement and cluster assignment. Different methods are evaluated on a public dataset (a derivative of the SINS dataset) used in the challenge on Detection and Classification of Acoustic Scenes and Events (DCASE) in 2018. Our method obtains the normalized mutual information (NMI) score of 54.46%, and the clustering accuracy (CA) score of 63.64%, and outperforms state-of-the-art methods in terms of NMI and CA. In addition, both computational complexity and memory requirement of our method is lower than that of previous deep-model-based methods.
Codes: https://github.com/vinceasvp/domestic-activity-clustering-from-audio

## I. INTRODUCTION

Automatic estimation of human activities is a key technique for safety monitoring of public space and smart home. Current works mainly include two classes: vision-based and wearable-sensor-based [1]. When adopted to estimate the domestic activities, these two classes of methods are deficient and even unsuitable due to practical reasons. For example, images and videos that are recorded at home are sensitive information, and cameras have blind areas and are impacted by illumination. In addition, people often forget or do not like to wear sensors.

To overcome the deficiencies of these two classes of methods above, this work focuses on the audio-based method. Microphones for acquiring audio clips can be easily deployed in various rooms instead of being mounted on human body, and are not affected by light and direction. A domestic activity from audio is considered as an acoustic scene [2], and contains one or some related sound events that appear in a location of the room, such as *Cooking*. Hence, the audio-based domestic activity estimation is related to the problem of acoustic scene classification (ASC).

ASC has been one of main tasks in some evaluation campaigns, such as the classification of events activities and relationships (CLEAR) [3], and the DCASE challenge [4]-[6]. In addition, many researchers independently carry out works on ASC [7]-[17]. The assumption of these works is that the labels of audio clips (including training data) are known beforehand. Hence, the main target of these works is to estimate pregiven class of acoustic scenes for testing audio clips. But, most audio clips are unlabeled due to the high labeling cost [18]. If the labels are unavailable, it is difficult even impossible to do ASC since labels are required for ASC. Conversely, clustering audio clips can be still done without labels. When labels are unknown, the problem we solve in this work is domestic activity clustering from audio clips.

The applications of domestic activity clustering mainly include discovering abnormal domestic activity, obtaining internal structure of unlabeled audio clips, and working as a pre-processing step of other tasks (e.g., audio clips after domestic activity clustering can be used to initialize deep models for ASC). Here, we take the discovery of abnormal domestic activity (e.g., falling) as an example for explaining the application of domestic activity clustering. The clusters of normal domestic activity (appearing frequently) contain a lot of audio clips, whereas the clusters of abnormal domestic activity (occurring rarely) have very few audio clips. In clustering process, the similarity between each new audio clip and each cluster center is used to determine whether this new audio clip belongs to one of the existing clusters or a new cluster. If a new cluster with very few audio clips occurs, the audio-based monitoring system will tell the management to check the elderly people's status.

Very few studies are conducted on acoustic scene clustering. The latest and most relevant work is done by Lin et al. [19] who propose a method of acoustic scene clustering by a convolutional capsule autoencoder network (CCAN). Their method is effective for domestic activity clustering, but the CCAN is a heavyweight network which mainly consists of standard convolution modules and complex capsule modules. Li et al. [20] propose a method of acoustic scene clustering method based on a randomly sketched sparse subspace, which is a shallow-model-based method by using mel-frequency cepstral coefficient as the input feature of a spectral clustering algorithm. Li et al. [18] present a method of acoustic scene clustering, where the input feature is the embedding learned

Corresponding author: Yanxiong Li.
This work was supported by international scientific research collaboration project of Guangdong Province, China (2021A0505030003), national natural science foundation of China (62111530145, 61771200), Guangdong basic and applied basic research foundation, China (2021A1515011454).

by a deep convolutional network and the clustering algorithm is the agglomerative hierarchical clustering. In addition, other deep models are adopted for clustering, such as stacked autoencoder (SA) [21], capsule autoencoder network (CaAN) [22], convolutional autoencoder network (CoAN) [23], long short-term memory network (LSTMN) [24]. In summary, previous works are either shallow-model-based method with hand-crafted features, or deep-model-based methods with heavyweight networks.

To further reduce the computational load and the size of deep model for lightweight applications, we propose a method of domestic activity clustering by a depthwise separable convolutional autoencoder network (DSCAN). The depthwise separable convolution (DSC) and the fully-connected layer are adopted in the DSCAN, and thus the computational complexity and memory requirement of our method can be reduced. In addition, the DSCAN is guided by a joint loss to co-optimize both embedding learning and cluster assignment and can obtain better clustering performance. Some methods are assessed on one public dataset. The results show that our method exceeds previous methods in terms of normalized mutual information (NMI) and clustering accuracy (CA), with lower computational complexity and memory requirement.

In summary, main contributions of this work are as follows. First, we propose a method of domestic activity clustering by the DSCAN. Second, we compare our method with state-of-the-art methods on a public dataset using multiple metrics.

## II. METHOD

In this work, we try to solve the problem of domestic activity clustering by clustering audio clips without using labels. The goal of this work is to correctly cluster audio clips that belong to the same class of domestic activity into one cluster. The proposed method of domestic activity clustering is a deep clustering method by the DSCAN. The framework of the DSCAN is shown in Fig. 1, which consists of an encoder, a decoder, and a clustering layer. Both the encoder and the decoder form the autoencoder which is used to learn embedding. The clustering layer is adopted for cluster assignment. A joint loss is designed to guide both embedding learning in the autoencoder and cluster assignment in the clustering layer. The encoder converts the input feature log-mel spectrum (log-mel) into an embedding, while the decoder reconstructs output feature log-mel' from the embedding.

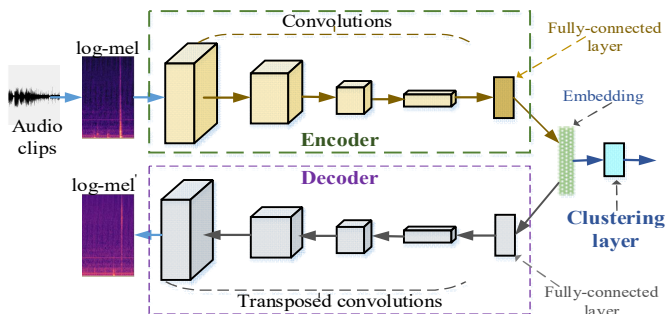

Fig. 1. The framework of the proposed DSCAN.

Compared to the deep models that are adopted in previous works [18], [19], [21]-[24], the proposed DSCAN has the following merits. First, inspired by the success of DSC [25], [26] for lightweight applications, the DSC instead of standard convolution is used in the encoder. The DSC can effectively reduce computational load and size of convolutional modules of the network [25]. Second, DSC-blocks and transposed Conv-blocks (Convolutional blocks) with residual structure (skip connection) are used in the DSCAN. The residual structure of the autoencoder can preserve information about the input feature maps and extract discriminative information from transformed feature maps in DSC-blocks. Third, a simple but effective fully-connected layer is proposed to replace the complex capsule module in the CaAN's and CCAN's encoder. The fully-connected layer can also reduce computational load and size of the DSCAN. In addition, the DSCAN is a deep model to jointly optimize both embedding learning and cluster assignment. Hence, the DSCAN is expected to be lightweight and can obtain better results compared to the deep models that are used in previous works.

It should be noted that both the DSC and the clustering layer are components of the proposed DSCAN. Although the DSC is introduced in [25], [26] and the clustering layer is added to an autoencoder in [23] for other tasks, the framework of the proposed DSCAN for domestic activity clustering is novel and is not adopted in previous works.

### A. Depthwise Separable Convolutional Autoencoder

The encoder and decoder of the autoencoder are denoted as $f(\cdot)$ and $g(\cdot)$, respectively. The autoencoder's goal is to minimize the mean squared error between its input and output over all input feature vectors, and is defined by

$$\min \frac{1}{N} \sum_{i=1}^{N} \left\| g(f(x_i)) - x_i \right\|_2^2, \quad (1)$$

where $N$ is the number of audio clips, and $x_i$ is the log-mel of the $i$th audio clip.

*Encoder*

DSC is the main module of the encoder, including two steps: depthwise convolution and $1 \times 1$ pointwise convolution [25]. The depthwise convolution is the key part for reducing the autoencoder's size. It applies one convolution filter to each input channel. The $1 \times 1$ pointwise convolution is used to combine the outputs of various depthwise convolutions, and determines the number of output channels.

The input feature maps, the output feature maps after convolution, and the convolutional kernel are denoted as $F_i \in \mathbb{R}^{w_i \times h_i \times c_i}$, $F_o \in \mathbb{R}^{w_i \times h_i \times c_o}$, and $K \in \mathbb{R}^{k \times k \times c_i \times c_o}$, respectively. $K \in \mathbb{R}^{k \times k \times c_i \times c_o}$, $K_d \in \mathbb{R}^{k \times k \times c_i \times 1}$, and $K_p \in \mathbb{R}^{1 \times 1 \times c_i \times c_o}$ denote the kernels of standard convolution, depthwise convolution, and pointwise convolution, respectively. Here, $w_i$, $h_i$, and $c_i$ (or $c_o$) are the feature-map's width, height, and the number of channels, respectively. $k \times k$ denotes the kernel size. The computational load and size of the network with standard convolution are $w_i \times h_i \times c_i \times c_o \times k \times k$, and $c_i \times k \times k \times c_o$, respectively [25]. The computational load and size of the network with DSC are $w_i \times h_i \times c_i \times (k^2 + c_o)$, and $c_i \times (k^2 + c_o)$, respectively [25]. If the $k \times k$ is set to $3 \times 3$, computational load of the network with DSC is theoretically 8 to 9 times less than that of the network with standard convolution.

As depicted in Fig. 2 (a), the encoder mainly consists of DSC-blocks. The elements of one DSC-block are depicted in Fig. 2 (b). As shown in Fig. 2 (a), the input feature log-mel is first processed by the operations of $5 \times 5$ standard convolution, batch normalization (BN), and ReLU. Then, the transformed feature maps are sequentially fed to five DSC-blocks. Finally, the output feature maps from DSC-block5 are processed by the operations of 1×1 standard convolution, BN, ReLU, and the fully-connected layer to output the embedding with 10 dimensions. The three digits in each bracket "{}" (e.g., {64, 78, 32}) from left to right denote the width of a feature-map (64), the height of a feature-map (78), and the number of channels (32), respectively.

As shown in Fig. 2 (b), the input feature maps are fed to a 1×1 convolution followed by the operations of BN and ReLU. Afterwards, the feature maps are transformed by the 3×3 depthwise convolution followed by the operations of BN and ReLU. Finally, a 1×1 pointwise convolution and the operations of BN and ReLU are conducted.

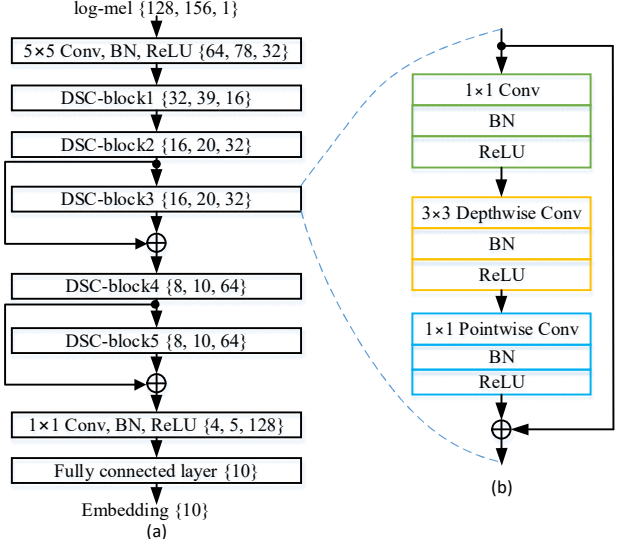

Fig. 2. (a) The encoder's structure; (b) One DSC-block's structure. Here, ⊕ denotes element-wise summation.

*Decoder*

Fig. 3 shows the decoder's structure which mainly consists of transposed Conv-blocks. Main module of the decoder is transposed Conv-block. The embedding is converted to a feature vector with dimension of 2560×1 by a fully-connected layer. Then, the feature vector is split into 20 vectors with dimension of 128×1, and five vectors with dimension of 128×1 are taken as a group to obtain four groups of such vectors, namely the reshaped feature map with dimension of 4×5×128. Next, the reshaped feature map is fed to the following transposed Conv-blocks in turn. Finally, the output feature log-mel' with the same shape of the log-mel is obtained after the 5×5 transposed convolution. Inspired by the success of residual structure in the ResNet [27], a skip connection is designed in each transposed Conv-block. Therefore, the information about the input embedding can be effectively transmitted to next transposed Conv-block.

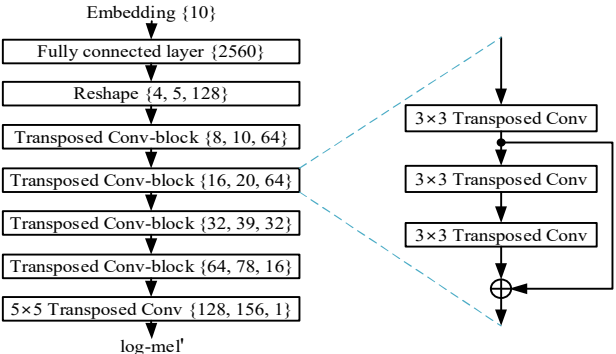

Fig. 3. The decoder's structure.

## B. Clustering Layer

The design of the clustering layer is inspired by [28], which takes the cluster center as trainable weight of the clustering layer. The training strategy in this work is like a form of self-training [29]. As in the self-training, an initial classifier and an unlabeled dataset are taken. Then, the dataset is labeled with the classifier for training the classifier on its own high-confidence predictions.

The student's $t$-distribution [30] for measuring the similarity between embedding $z_i$ and cluster center $u_j$ is defined by

$$q_{ij} = \frac{\left(1 + \alpha^{-1}\|z_i - u_j\|^2\right)^{-\frac{\alpha+1}{2}}}{\sum_{j'}\left(1 + \alpha^{-1}\|z_i - u_{j'}\|^2\right)^{-\frac{\alpha+1}{2}}}, \quad (2)$$

where $z_i=f(x_i)$, $\alpha$ is the degree of freedom of student's $t$-distribution and is set to 1 here, and $u_j$ is the mean vector of embeddings in cluster $j$ which is obtained by conducting the $K$-means clustering algorithm on embeddings. When the current mean vectors are not updated, the $K$-means clustering is terminated and thus the mean vectors of embeddings are obtained. $q_{ij}$ can be considered as the probability that embedding $z_i$ belongs to cluster $j$. As done in [28], the predefined target probability $p_{ij}$ that embedding $z_i$ belongs to cluster $j$, is defined by

$$p_{ij} = \frac{q_{ij}^2 / \sum_i q_{ij}}{\sum_{j'}\left(q_{ij'}^2 / \sum_i q_{ij'}\right)}. \quad (3)$$

With the help of the predefined target probability in Eq. (3), the clusters can be iteratively refined by learning from their high-confidence assignments.

## C. Joint Loss

A joint loss is defined to jointly optimize both embedding learning (realized by the autoencoder) and cluster assignment (realized by the clustering layer). The reconstruction loss for embedding learning is defined by

$$L_r = \frac{1}{N}\sum_{i=1}^{N}\|g(f(x_i)) - x_i\|_2^2, \quad (4)$$

where $N$ is the number of audio clips, and $x_i$ is the log-mel of the $i$th audio clip. The clustering loss $L_c$ is defined as a Kullback-Leibler (KL) divergence [28], [31] between the distribution of soft labels $Q = \{q_i\}$ and the predefined target distribution $P=\{p_i\}$. $L_c$ is computed by

$$L_c = KL(P \| Q) = \sum_i \sum_j p_{ij} \log \frac{p_{ij}}{q_{ij}}, \quad (5)$$

where $q_{ij}$ is first computed by Eq. (2), and then $p_{ij}$ is computed by Eq (3) in order to obtain $L_c$.

The joint loss $L_J$ is defined by

$$L_J = L_r + \beta L_c, \quad (6)$$

where $\beta > 0$, is a coefficient for balancing the contribution of $L_r$ and $L_c$ to the value of $L_J$. It should be noted that adding $L_c$ to $L_r$ will affect the feature reconstruction of autoencoder. The greater the proportion of $L_c$ in the joint loss $L_J$, the greater the influence of $L_c$ on feature reconstruction and thus the more the embedding space is distorted. When the value of $\beta$ is set properly, the distortion of the embedding space caused by the addition of $L_c$ will be within an acceptable range, and meanwhile the clustering loss can be optimized for obtaining more satisfactory clustering results.

The DSCAN's optimization guided by $L_J$ is summarized in Table I. After updating the DSCAN under the guidance of $L_J$, the optimized clustering results are obtained.

TABLE I. THE OPTIMIZATION OF THE DSCAN GUIDED BY $L_J$

| Initialization: |
|---|
| ① Pretrain autoencoder with $\beta=0$ (i.e., $L_J = L_r$) to get target distribution; |
| ② Initialize cluster centers by $K$-means algorithm on embeddings; |
| ③ Set $\beta$ to be a fixed non-zero value. |
| **Repeat:** |
| ① Update autoencoder's weights and cluster centers by backpropagation and Adam algorithms [32]; |
| ② Update predicted probability $q_{ij}$ by (2) and target probability $p_{ij}$ by (3). |
| **Until** the change of cluster label assignments between two adjacent iterations is less than a threshold $\varepsilon$ or reaching the maximum iteration. |
| **Output:** Optimized results of domestic activity clustering. |

## III. EXPERIMENTS

This section first describes experimental data and setup, and then presents results and discussions.

### A. Experimental Data

To our best knowledge, there is only one public audio dataset of domestic activities, namely the dataset of Task 5 of DCASE-2018 challenge, which is a derivative of the SINS dataset [33] and consists of audio clips of 10 seconds. Each audio clip contains one domestic activity. Table II lists detailed information about experimental data that include 9 daily domestic activities. All audio clips in Table II are used as testing data for clustering. During the clustering process, the labels of these audio clips are not used.

TABLE II. DETAILED INFORMATION ABOUT EXPERIMENTAL DATA

| Activities | No. of clips | Activities | No. of clips |
|---|---|---|---|
| Dishwashing | 1424 | Vacuum cleaning | 972 |
| Cooking | 5124 | Social activity | 4944 |
| Absence | 18860 | Watching TV | 18648 |
| Eating | 2308 | Working | 18644 |
| Other | 2060 | **Total** | **72984** |

To experimentally set the proper value of $\beta$ in our method, we generate a development dataset by selecting audio clips from the development dataset of Task 1A of DCASE 2019 [34]. The development dataset contains 9 classes of acoustic scenes, including *Airport*, *Shopping mall*, *Metro station*, *Pedestrian street*, *Public square*, *Street traffic*, *Bus*, *Metro*, and *Urban park*. Each class of acoustic scene includes 1440 audio clips, and thus the total number of audio clips in the development dataset is 12960. The duration of each audio clip is approximately 10 seconds.

### B. Experimental Setup

All experiments are implemented on a machine: one Intel CPU I7-6850K with 3.6 GHz, one RAM of 128 GB, and four NVIDIA 1080TI GPUs. All methods are implemented by the toolkits of Keras, TensorFlow and scikit-learn.

CA and NMI are two common metrics used for clustering [35], whose definitions are presented below. It is assumed that $n_{ij}$ represents total number of audio clips in cluster $i$ that belong to domestic activity $j$. $N_g$ and $N_c$ are total number of classes of domestic activities (the real number of clusters) and total number of clusters (the predicted number of clusters), respectively. $N_s$ denotes total number of audio clips. $n_{\cdot j}$ and $n_{i\cdot}$ stand for total number of audio clips of domestic activity $j$ and total number of audio clips in cluster $i$, respectively. The three formulas in Eq. (7) establish relationships among the variables above:

$$n_{i\cdot} = \sum_{j=1}^{N_g} n_{ij} \quad , \quad n_{\cdot j} = \sum_{i=1}^{N_c} n_{ij} \quad , \quad N_s = \sum_{i=1}^{N_c} \sum_{j=1}^{N_g} n_{ij} \quad . \quad (7)$$

NMI and CA are adopted for measuring the agreement between the predicted clusters and the ground-truth classes. NMI is defined by

$$NMI = \frac{\sum_{i=1}^{N_c} \sum_{j=1}^{N_g} n_{ij} \log\left(\frac{N_s \times n_{ij}}{n_{i\cdot} \times n_{\cdot j}}\right)}{\sqrt{\left(\sum_i n_{i\cdot} \log \frac{n_{i\cdot}}{N_s}\right)\left(\sum_j n_{\cdot j} \log \frac{n_{\cdot j}}{N_s}\right)}} \quad . \quad (8)$$

NMI is equal to 1 if the predicted clusters perfectly match the ground-truth classes. On the contrary, NMI is close to 0 if the audio clips are partitioned randomly. CA is defined by

$$CA = \frac{1}{N_s}\left[\sum_{k=1}^{N_s} \delta(y_k, \text{map}(c_k))\right] \quad , \quad (9)$$

where $c_k$ and $y_k$ represent the predicted and true cluster labels, respectively, of the $k^{th}$ audio clip. If $y = c$, $\delta(y, c)$ is equal to 1, otherwise $\delta(y, c)$ is equal to 0. map(·) is a permutation function which maps each cluster label to a ground-truth label. Based on their definitions above, it can be known that NMI is an information-theoretic measure of the clustering quality and CA is a permutation-mapping measure of the clustering quality. The higher their values are, the higher the clustering quality is.

The values of main parameters of our method are given in Table III. They are either often used in previous works or common empirical values. In addition, the value of $\beta$ has direct impact on the performance of our method, whose settings will be discussed in next subsection.

TABLE III. THE SETTINGS OF MAIN PARAMETERS OF OUR METHOD

| Type | Parameter's settings |
|---|---|
| log-mel | Frame length/overlap: 128ms/64ms<br>Dimension of log-mel: 128 |
| DSCAN | Number of pretraining iterations: 200<br>Number of maximum iterations: 4000<br>Batch size: 32<br>Learning rate: 0.001<br>Dimension of embeddings: 10<br>Threshold $\varepsilon$: 0.05<br>Neuron number of fully-connected layer: 2560<br>Number of cluster centers: 9 |

### C. Results and Discussions

First, we carry out one experiment on the development dataset for setting the proper value of $\beta$ in our method. The values of $\beta$ are tuned from 0.1 to 0.9. As shown in Fig. 4, when the value of $\beta$ is equal to 0.3, our method obtains the highest scores of both CA and NMI on the development dataset. However, when the values of $\beta$ deviate from 0.3, the scores of both CA and NMI steadily decrease. Therefore, the proper value of $\beta$ is set to 0.3 when our method is evaluated on the testing dataset.

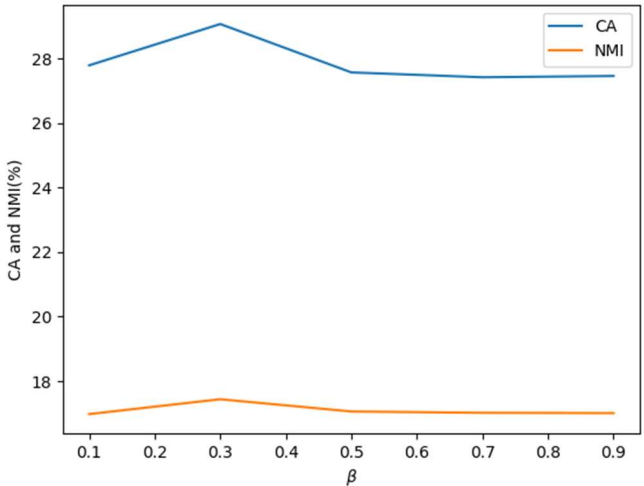

Fig. 4 The impact of $\beta$ on the performance of our method on the development dataset.

Then, we compare our method to five deep-model-based methods, including: the SA-based [21], CaAN-based [22], CoAN-based [23], LSTMN-based [24], and CCAN-based [19]. Main parameters of the previous methods are set according to the suggestions in corresponding references. Experimental results of the previous methods are obtained by implementing the previous methods by ourselves. The labels of audio clips are not adopted in all methods during clustering procedure, but they are used for performance evaluation. All experimental data listed in Table II are adopted as testing data for clustering. Experimental results obtained by different methods are given in Table IV.

TABLE IV. PERFORMANCE COMPARISON OF DIFFERENT METHODS

| Methods | CA (%) | NMI (%) | MACs (K) | MS (K) |
| --- | --- | --- | --- | --- |
| SA-based [21] | 45.47 | 38.48 | 2576.5 | 1842.9 |
| CaAN-based [22] | 54.81 | 46.03 | 3054.3 | 632.6 |
| CoAN-based [23] | 49.62 | 42.60 | 2099.4 | 118.6 |
| LSTMN-based [24] | 50.36 | 43.08 | 1158.5 | 1004.4 |
| CCAN-based [19] | 61.91 | 53.84 | 1451.9 | 725.3 |
| **DSCAN-based** | **63.64** | **54.46** | **1048.8** | **72.4** |

As shown in Table IV, our method obtains CA score of 63.64% and achieves absolute gains by 18.17%, 8.83%, 14.02%, 13.28%, and 1.73% over the methods of SA-based, CaAN-based, CoAN-based, LSTMN-based, and CCAN-based, respectively. As for the score of NMI, our method produces 54.46%, and achieves absolute gains by 15.98%, 8.43%, 11.86%, 11.38%, and 0.62% over the methods of SA-based, CaAN-based, CoAN-based, LSTMN-based, and CCAN-based, respectively. In summary, our method obtains higher scores of CA and NMI than previous methods.

In addition, our method is compared to previous methods in terms of computational complexity using the metric of Multiply-Accumulate operations (MACs) and in terms of memory requirement using the metric of Model Size (MS). MACs denotes the number of multiplication and addition operations of a network. MS represents the number of parameters of a network. The lower the values of MACs and MS are, the lower computational complexity and memory requirement of the method are. The values of MACs and MS of different methods are listed in Table IV. In terms of computational complexity, the MACs of our method is 1048.8 K which is lower than that of other deep-model-based methods. In terms of memory requirement, the MS of our method is 72.4 K which is significantly less than that of the previous methods.

Based on the results above, we can conclude that our method exceeds previous methods in terms of both NMI and CA, and has advantage over previous methods in terms of both computational complexity and memory requirement. The possible reasons are as follows. First, the residual structures adopted in the DSC-blocks and transposed Conv-blocks can preserve information about input feature maps, and meanwhile can extract discriminative information from the transformed feature maps. Hence, the embedding learned by the DSCAN can effectively represent differences of time-frequency properties among different domestic activities and obtains better results for domestic activity clustering. Second, both the fully-connected layer (instead of a complex capsule module) and the DSC (instead of standard convolution) modules are used in the DSCAN. Hence, the DSCAN-based method is with lower computational complexity and lower memory requirement compared to the previous methods.

To visually show the results of our method, we use the t-SNE [30] to map embeddings into two-dimensional space and thus obtain spatial distribution of various clusters as depicted in Fig. 5. The Python libraries of *scikit-learn* and *matplotlib* are used to reduce the dimensionality of embeddings and plot Fig. 5, respectively. Though most audio clips of the same class are merged to their respective cluster centers, there are confusions among different clusters. For example, audio clips of *Other* are scattered to other clusters (e.g., *Absence, Eating*). The reasons for the confusions are probably as follows. Differences of time-frequency properties of these domestic activities are not effectively represented, and there are overlapping regions in the distributions of their embeddings.

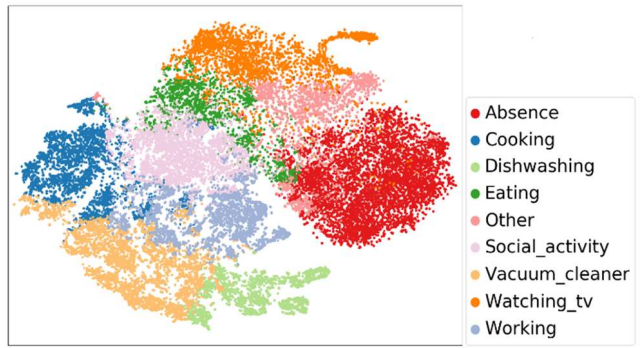

Fig. 5. Visualization of clustering results obtained by our method.

IV. CONCLUSIONS

In this paper, we tackled the problem of domestic activity clustering from audio clips using the proposed DSCAN. Our method outperforms previous methods in terms of CA and NMI. In addition, its computational complexity and memory requirement are lower than that of the previous methods. However, the problem of domestic activity clustering is still challenging due to the following causes. First, the methods of domestic activity clustering work in an unsupervised way without using labels. Second, there are unbalances of data amount of audio clips for different classes. Third, there are overlapping regions in the feature distributions of audio clips for various domestic activities.

In the future work, we plan to deploy our method in the terminals with low-computing resources. In addition, we will estimate activities from audio clips in other situations, such as roads, train/metro stations.